\documentstyle[12pt]{article}
\newcommand{\cmthree}{cm$^{-3}$}

\newcommand{\kms}{km\,s$^{-1}$}       
\newcommand{\um}{$\mu$m}                                 
\newcommand{\molh}{H$_{2}$}                              

\newcommand{\msun}{${\rm M}_{\odot}$}

\newcommand{\gapprox}{$\stackrel {>}{_{\sim}}$}   
\newcommand{\lapprox}{$\stackrel {<}{_{\sim}}$}
\newcommand{\about}{$\sim$}                       
\newcommand{\adeg}{$^{\circ}$}

\begin{document} 

\begin{center}
\Huge
{\bf First Life in the Universe}

\vspace*{1.25cm}
\Large
{\bf Ren\'e Liseau}

\vspace*{1.0cm}
\normalsize
Department of Space, Earth and Environment, Chalmers University of Technology, Onsala Space Observatory, SE-439 92 Onsala, Sweden, {\footnotesize\texttt{e-mail: ${\rm rene.liseau@chalmers.se}$}}

\end {center}
\vspace*{1.5cm}

\noindent
This presentation is aimed at first year students of physics, but also at the general, science interested, public. It is therefore deliberately kept free from technical details. These can be found in the references provided in footnotes. 

The paper is organised in several chapters, some of which might be skipped by the knowledgeable reader.\\

\begin{center}
INDEX
\end{center}

\noindent
CHAPTER\hfill PAGE \\
\\
Short Summary \hfill 3 \\
\\
1. What this is all about \hfill 3 \\
\\
2. Things we call life  \hfill 4\\
\\
\hspace*{0.4cm}2.1 What is life made of? \hfill 4\\
\\
\hspace*{0.8cm}2.1.1 Is life restricted to the matter world? \hfill 4\\
\\
\hspace*{0.8cm}2.1.2 Could it be possible that there exists life made of antimatter? \hfill 4\\
\\
\hspace*{0.8cm}2.1.3 Could it be possible that there is life in the dark matter world? \hfill 4 \\
\\
\hspace*{0.4cm}2.2 How do we recognise living things?  \hfill 5\\
\\
\hspace*{0.4cm}2.3 What is life doing?  \hfill 5\\
\\
\hspace*{0.4cm}2.4 What is the structural architecture of life?  \hfill 5\\
\\
3. How it all started  \hfill 5\\
\\
\hspace*{0.4cm}3.1 What is the age of the universe?  \hfill 5\\
\\
\hspace*{0.4cm}3.2 The first elements  \hfill 6\\
\\
\hspace*{0.4cm}3.3 The first stars  \hfill 7\\
\\
\hspace*{0.8cm}3.3.1 Population III: low mass stars  \hfill 7\\
\\
\hspace*{0.8cm}3.3.2 Population III: high mass stars  \hfill 8\\
\\
\hspace*{0.4cm}3.4 The Dark Ages of the universe  \hfill 9\\
\\
\hspace*{0.8cm}3.4.1 The standard model  \hfill 9\\
\\
\hspace*{0.8cm}3.4.2 How dark were the Dark Ages and for how long?  \hfill 10\\
\\
4. Coming back to the issue of life  \hfill 11\\
\\
5. Possibilities to detect extraterrestrial life  \hfill 12\\
\\ 
\hspace*{0.4cm}5.1 The nearby universe  \hfill 12\\
\\
\hspace*{0.8cm}5.1.1 Solar system exploration  \hfill 13\\
\\
\hspace*{0.8cm}5.1.2 Stars and  planets and the ISM in the Milky Way  \hfill 13\\
\\
\hspace*{0.4cm}5.2 The distant universe  \hfill 14\\
\\
6. Conclusions  \hfill 14\\
\\
7. Resum\'e \hfill 15 \\
\\
Bibliography  \hfill 16\\
\\
Appendix  \hfill 19

%

\newpage

\noindent
{\small{\bf Short Summary.}} 
Here, we ask a simple question, i.e. ``at what cosmic time, at the earliest, did life first appear in the universe?"  Given what we know about the universe today, there may be some partial answers to this question, but much will still have to be left to speculation. If life in general requires stars as its primary energy source and uses elemental building blocks heavier than those initially produced in a Big Bang scenario, first life could have appeared, when the universe was considerably less than 0.1 billion years old. At that time, heavy element producing hypernovae exploded at corresponding redshifts\footnote{The observed wavelength of a spectral line can produce a shift $z = (\lambda_{\rm obs}-\lambda_{\rm lab}) / \lambda_{\rm lab}$. If the observed wavelength is larger than that measured in the lab, the shift is toward the red of the electromagnetic spectrum. $z$ is a directly measured quantity and as such independent of any model.} $z$\,\gapprox\,45, significantly higher than commonly assumed ($z \approx 6$). The recent discovery of a galaxy at $z > 9$ could provide supporting evidence for the hypothesis of a very much shorter time scale than what is widely believed. \\

\section{What this is all about}

One of the questions that I have been wondering about for a long time is ``When, at the very earliest, could life have emerged in the universe? And was that a unique event or did it happen several times, over and over again?" 

One is generally thinking of life that originated and evolved on Earth. But, could there have been occasions in other places and at other times for this to happen? In other words, at what stage, in the history of the universe, were the {\it absolutely necessary} ingredients abound in sufficient abundance and the overall conditions acceptable for life to get started? A less restrictive question would be ``when were the necessary building blocks available" and life would have had the possibility to develop? The latter question is often used by geocentric-oriented people, with the proviso ``life as we know it".

However, this presents us immediately with a problem, leading us into a blind all\'e: we don't know the {\it life as we know it}, since we, in spite of numerous serious attempts to do so, have not found a way to unambiguously defining it nor to making it in the lab\footnote{See, e.g., the review by Bich, L. \& Green, S. (2018, Synthese 195, 3919) {\it Is defining life pointless? Operational definitions at the frontiers of biology}}. 

One of the major difficulties to really {\it know life} in a more general sense stems from the fact that we are dealing with the statistics of one - we are aware of its existence only on Earth. The statistics of one is a contradiction in terms and of course nonsense. But it illustrates the important and well-known fact that we have nothing else to compare with and to learn from, opening the field to enormous speculation. 

Most often, discussions about life are hampered already at the very start, depending on what axioms one is equipped with. For instance, a teleological standpoint, based on personal belief and conviction, excludes a meaningful communication with other views. Often, the axiom is a divine plan, according to which humans are the crown of creation, fulfilling the plan. With this axiom, explanations for ``why?" and ``what is the plan?" need not to be offered by its defenders and non-fitting ideas need not to be considered.

Another phalanx of debaters represents people that are expressing essentially the opposite view, i.e. life is a number of random processes without external intervention, involving more or less coincidental accidents. These can lead to the idea of a darwinistic kind of evolution, the outcome of which is not pre-determined and, hence, not foreseeable.

Here, we will take a rather pragmatic, but non-the-less scientific, approach requiring that a logical line of thought called {\it theory} has to be able to withstand testing through repeated experiments or observations. We will try to argue free from prejudice. At the same time, we also realise that it will likely become unavoidable not to retrieve examples and arguments based on life on Earth. Is it then at all meaningful to continue talking about life? I think, yes, since it may very well be of interest to the general community to scrutinise the matter out of the box and in some depth. 

\section{Things we call life}

First, we need, however, agree on a few assumptions. Even though one has not been able to provide a satisfactory definition of life, we may still ask 

\subsection{What is life made of? }

Without supporting and scientifically verifiable evidence for life to exist in an immaterial, spiritual world, we can assume that life belongs to the material world. We may then ask

\subsubsection{Is life restricted to the matter world?}

We are aware of life forms in only one manifestation of ordinary matter. That does not necessarily mean that life is entirely bound to matter we are generally familiar with. A slightly irritating fact might be that ordinary matter, like us and the stars, is not by far the most abundant phase state of gravitating stuff. 

\subsubsection{Could it be possible that there exists life made of antimatter?}

Antimatter has the same physical properties as matter, but with opposite electric charge and, if appropriate, opposite baryon number, or opposite spin. An example is antihydrogen\footnote{Oelert W., for the PS210-Collaboration (1997), Nuc.Phys. B, Proceedings Suppl., 56, 319: {\it Production of anti-hydrogen at CERN}}, i.e. an antielectron (called positron) bound to an antiproton. In addition, some particles are their own matter and antimatter, such as the photon and some other bosons. 
 
In principle, one cannot see any plausible reason, why there should not exist any ``antilife" somewhere in the universe. We should avoid to shake hands with any hypothetical antihumans, though. When matter and antimatter meet and bounce into each other, they completely annihilate into high-energy radiation ($\gamma$-ray photons).

However, the chances to encounter or even detect antilife may be rather slim, as the observable universe consists of an overwhelmingly larger amount of matter, with antimatter being highly elusive. Until this day, this matter-antimatter imbalance has remained one of the biggest mysteries of the universe.

\subsubsection{Is there life in the dark matter world?} 

The question, whether life could conceivably also has manifested itself in the {\it dark matter} (DM) world appears highly academic. However, the way by which DM has been defined would make it highly unsuitable for chemical reactions, as it does not interact in any other way than gravitationally\footnote{Possible interactions between baryonic and dark matter have been speculated about by Bowman et al., 2018, Nat., 555, 67, {\it An absorption profile centred at 78 megahertz in the sky-averaged spectrum}, and references therein.}. However, in contrast to antimatter, we do not have a clue what DM is, in spite of the fact that dark matter stands apparently for about 85\% of all gravitating matter in the universe. So, we need to leave this question open.

\subsection{How do we recognize living things?} 

We all have a clear idea about what non-living things are, including those that once were alive but now are dead. We have an {\it instinctive} understanding of what life is, but have difficulties to phrase this insight into strictly defining words. Maybe, one way to go about it, is to take an operational path, asking a different question, viz. 

\subsection{What is life doing?} 

Living individuals eat, digest and excrete, i.e., they metabolize. However, this would also be applicable to a car engine, which we would not call a living thing: it is made of ordinary matter, consumes fuel, transforms that nutrition to energy in the form of heat and expels gases through its exhaust. And, like life, it ages and in the end, it dies.

It is clear, therefore that the points made above are necessary, but obviously not sufficient, conditions for life. However, living organisms do also something else, i.e., they build self-replicating entities and produce offsprings. A car engine does not do that, but on the other hand, crystals show growth patterns and make copies of themselves. They too, we feel, are non-living things. In summary, the doings of life seem not to be unique.

\subsection{What is the structural architecture of life?}

It is reasonable to assume that life follows universal recipes and patterns that have been applied throughout the history of the universe, i.e. the building from smaller to larger things in a hierarchical fashion, from quarks to nucleons to atoms to molecules and so on, up to superclusters of galaxies. On Earth too, life forms developed from the simplest to more complex aggregates of matter. Made of biomolecules, these display an interesting (and unexplained) feature: as building blocks, life on Earth uses exclusively homochiral molecules, i.e. molecules with the same ``handedness" (chirality)\footnote {e.g., L (left-handed) for amino acids and D (right-handed) for sugars; see articles in: Lough, W. J., \& Wainer, I. W., eds., 2002, {\it Chirality in natural and applied science}, Oxford: Blackwell Science.}. 

In addition, all life on Earth seems confined, in cell-like ``bags", rather than being distributed, like in a gas or fluid\footnote{Hypothetical alternatives can be found in, e.g., the books ``The Black Cloud", by Fred Hoyle (1957) or ``Solaris" by Stanislaw Lem (1961).}. It is not clear, to what extent this is a general property of life. It is also not clear, whether life always needs a planet or whether there exist(ed) life forms in the vast volumes of space in between the stars, in the so called interstellar medium (ISM).

\section{How it all started}
 
\subsection{What is the age of the universe?} 

When addressing the physicochemical possibility for the first appearance of life, we need to know when to start the clock. A first rough estimate would be given by the inverse of the Hubble constant, $H_0$, which measures the current expansion of the universe, and where $1/H_0$ has the units of time. The actual value of $H_0$ has become a hotly debated issue. Based on data referring to the early\footnote{From the analysis of data from ESA«s Planck mission the value of $H_0=67.4 \pm 0.5$\,\kms\,Mpc$^{-1}$ has been derived (The Planck collaboration, 2019,  arXiv:1807.06209v2, {\it Planck 2018 results. VI. Cosmological parameters}; ESA stands for European Space Agency). This results in a Hubble time $1/H_0 = 14.5 \pm 0.1$\,Gyr. The Planck result has recently been confirmed by an independent, ground based experiment (Aiola, S., et al. 2020, arXiv:2007.07288, {\it The Atacama Cosmology Telescope: DR4 Maps and Cosmological Parameters}) yielding $H_0 = 67.6 \pm 1.1$\,\kms\,Mpc$^{-1}$. The CMB (Cosmic Microwave Background) is the relict radiation from the hot Big Bang at the beginning of the universe. The photons of the CMB we receive today originated when the universe was about 470 thousand years old.} and the late\footnote{Observations of standard candles (Cepheids and Supernovae type Ia, Riess et al. 2019, ApJ 876, 85) on one hand and of gravitational lenses on the other (Wong et al. 2020, MNRAS, arXiv:1907.04869v2) in the local universe ($z$\,\lapprox\,2) have resulted in the combined value $H_0=73.8 \pm 1.1$\,\kms\,Mpc$^{-1}$, i.e. $1/H_0 = 13.2 \pm 0.2$\,Gyr.} universe, two values have been found that differ by more than a billion years, viz. $1/H_0 = 14.5$\,Gyr and $13.2$\,Gyr, respectively. A billion years is a long time, even by astronomical standards. With regard to life on Earth, a lot of evolution happened during a billion years. Also for that reason would we like to know better the true age of the universe.

The two determinations of $H_0$ are based on three studies that are exploiting completely different methods and are entirely independent of each other. To determine the actual age of the universe requires the use of a particular theoretical, or cosmological, model. In fact, this discrepancy in the values of $H_0$ may have profound implications for our understanding of the universe, and hence require the development of complementary or even alternate cosmologies\footnote{Bernal J.L., Verde L. \& Riess A.G. (2016, JCAP 019, 1) discuss {\it The trouble with $H_0$} at considerable depth. Since the appearance of their article, the discrepancy has become even more severe: Riess A.G. 2019, Nat. Rev. Phys., {\it The expansion of the Universe is faster than expected.} Current ideas of how to resolve the $H_0$ issue is to introduce further ad hoc assumptions into the existing models.}. 

However, for the currently widely accepted $\Lambda$CDM cosmology ($\Lambda$ Cold Dark Matter),  one derives an age\footnote{For the redshift-age relation, $\tau(\Omega_{\rm m}, \Omega_{\Lambda}; z)$, we adopt the integrated Friedman equation by Thomas \& Kantowski (2000, Phys. Rev. D 62, 103507, Eq.\,18), i.e. valid for a Friedman-Lema\^{i}tre-Robertson-Walker metric with $\Omega_0 = \Omega_{\Lambda} + \Omega_{\rm m} = 1$, $ \Omega_{\Lambda}  \ne 0$ and $\Omega_{\kappa} = 0$. The last term describes the geometrical flatness of the universe, the value of which, on the basis of Planck data, has been challenged in an article by Di Valentino E., Melchiorri A. \& Silk J. (2020, Nature Astronomy, 4, 196).} of 13.8\,Gyr in the case of the ``early time", and of 12.6\,Gyr in the case of the ``late time" measurement\footnote{The CMB stems from the time when matter and photons decoupled, at redshift $z \approx 1100$; in contrast, the light from Ia SNe can be measured with confidence in galaxies out to redshift $z \approx 2$; e.g. Smith et al. 2018, ApJ 854, 37.}. Obviously, at most one of these numbers can be correct. The younger age seems to be in conflict with ages that have been quoted in the literature for the oldest stars in the Galaxy\footnote{See, e.g., Howes et al. 2015, Nat. 527, 484 and ${\rm https\!\!:\!\!//en.wikipedia.org/wiki/List\_of\_oldest\_stars}$}. 

As our interests lie in the early phases of the universe, throughout this paper and unless otherwise stated, the value of $H_0$ derived by the Planck-collaboration will be used.

\subsection{The first elements}

The success of the Big Bang model has largely been based on its  feat to correctly reproduce observed abundances of the elemental species\footnote{and to predict (R.\,Alpher, 1948, arXiv:1411.0172v1) the existence of the isotropic background radiation (CMB), that was serendipitously discovered in 1964 by Penzias and Wilson (1965, ApJ 142, 419). The discovery was theoretically explained by Dicke et al. (1965, ApJ 142, 414) in the same journal.}. In particular, the Big Bang synthesized merely the lightest nuclei with the number of protons and neutrons $A \le 7$, i.e. H, He, Li and their isotopes. After these first few minutes ($\approx 20\,\mu$yr) of the explosion, the extremely rapid expansion resulted in the dramatic decrease of the density and the temperature to drop to far below ten billion degrees\footnote{In the early, radiation dominated, universe, the temperature drops approximately as the inverse of the scale factor, $T \propto 1/a$, hence as $t^{-0.5}$. The scale factor is related to the Hubble parameter $H(t) = \dot a/a$, with the definition $a_0=1$ for $H(t)=H_0$.}. This prevented the nucleosynthesis of heavier species, the accomplishment of which would need energetic nuclear burning inside stars.

 \subsection{The first stars}

After the initial Big Bang nucleosynthesis was completed, most of the matter in the universe was in the form of hydrogen nuclei, with slightly less than a quarter of the mass being in the form of helium nuclei\footnote{The standard model primordial value of the helium abundance is $Y_{\rm p} = 0.24672 \pm 0.00017$ (e.g., Cooke \& Fumagalli, 2018, Nature Astronomy 2, 957, and references therein).}. In addition, the remaining species accounted for another small fraction.

Stars with masses around that of the sun (\msun) convert hydrogen to helium by predominantly the so called proton-proton cycle. In the pp-cycle, four hydrogen nuclei combine into one helium nucleus ($4\,p \rightarrow 1\,\alpha$). Above about 1.3\,\msun, the so called CNO cycle\footnote{CNO stands for Carbon C, nitrogen N and oxygen O. These act merely as catalysts and do not contribute to the nuclear energy generation.} dominates the helium production, but also this process converts four H-nuclei into one He-nucleus. In all fusion processes, a number of other particles are also produced that carry away some part of the generated energy. The total energy released through fusion is the binding energy difference of the involved species. 

The great majority of the observed stars are stars that produce most of their energy through fusion of hydrogen. Because of that, their basic physics is the same and apply to all of them and, consequently, they constitute what is called the main sequence (MS) of stars\footnote{When plotting the stellar surface temperature against the radiated power, the data lie in a relatively narrow band across this so called Hertzsprung-Russell diagram (HR diagram).}. They are stable and change only relatively slowly, producing energy over considerable amounts of time. However, their exact lifetimes are critically depending on their mass, with stars of lower mass living longer. 

After the MS phase, the fusion of helium nuclei yields primarily carbon via the triple-alpha process ($3\,\alpha \rightarrow 1\,^{12}$C). By fusing with He-nuclei, the $\alpha$-process proceeds after carbon to produce $^{16}$O, $^{20}$Ne and so on, i.e. predominantly nuclei consisting of multiples of four with even numbers of protons.

\subsubsection{Population III: low mass stars}
 
In roundish numbers, the MS-life of the sun is 10 billion years. To order of magnitude, the ages of stars of various masses can be obtained from $t_{\rm MS} = 10^{10}/M^{\rm {m}}$\,yr, where m=3 for solar masses $M=0.1$ to 3. The life expectancy for a star only a tenth as massive as the sun ($M=0.1$) is thus ten thousand billion years, a thousand times the age of the universe\,!  In contrast, a star only three times as massive as the sun lives for only 370 million years\footnote{To put this into perspective:  370 million years ago, the first sharks appeared on Earth  (Aidan, 2012, http://www.elasmo-research.org/education/evolution/ancient.htm).}. For still higher masses, the time dependence becomes less steep, so that, e.g., m$(10)=2.5$ and m$(\ge 30)=2$, resulting in MS-lifetimes, respectively, of about 30 million and less than 10 million years. The latter is comparable to the evolutionary history of mankind.

It becomes evident, that stars lighter than the sun and which were born at the earliest times of the universe, should still be around, lingering on the main sequence. In favorable cases, some of them may be directly observable by us and provide us with valuable information regarding the early history of the universe. 

Since these stars predominantly burned hydrogen according to the pp-cycle, heavier elements than helium should essentially not show up in these stars (zero-metallicity stars), or at least be highly depleted. Such stars are called ultra-metal-poor, with the record holder having an iron abundance a million times less than the sun\footnote{[Fe/H]$\,= -6.2 \pm 0.2$, for an 1d LTE model; Nordlander et al., 2019, MNRAS 488, L109.  Aside from iron, the spectrum of this particular star displays also trace amounts of other heavy elements (C, Ca, Mg and Ti), indicating that these were formed in a supernova of stellar progenitor mass of \about10\,\msun.}. Concepts become here a little fuzzy and some astronomers term these stars pop\,III, others reserve this epithet for the truly very first stars. So far, searches for zero-metallicity stars in the Galaxy have been unsuccessful and fruitless. 

\subsubsection{Population III: high mass stars} 
 
On the main sequence, hydrogen fusion results in the build-up of a helium core at the center of the stars. After the MS life, stars evolve rather rapidly and expel, by means of stellar winds or supernova explosions (stars heavier than about 8\,\msun\ will end their lives as supernovae), their nuclear ashes into space. That material then can give rise to new generations of stars, which are made of progressively more enriched elements. The sun, for instance, being born nearly ten billion years after the Big Bang, shows elemental abundances that are much higher than the primordial ones.
 
 According to the theory of stellar evolution, stars with masses larger than the sun are generating their thermonuclear energy running the CNO-cycle. In order to ignite the initial steps of this process, appreciable amounts of carbon as catalytic and nuclear fuel are needed. In the very early universe, however and since carbon is produced only after the main sequence phase, stars of masses lower than a certain mass will not have had time to evolve off the MS. 

Following the Big Bang nucleosynthesis, stars that are made of only hydrogen and helium are difficult to form. In order to accomplish that, stars have to be extremely heavy, maybe up to 1000\,\msun, which is much larger than the highest masses of currently observed stars ($<100$\,\msun). 

The evolution of Pop\,III (``zero metallicity") stars of several hundred solar masses has been computed using detailed theoretical models\footnote{Baraffe, Heger \& Woosley, 2001, ApJ 550, 890}. For masses between 120 and 500\,\msun, these models yield MS lifetimes between 2.35 and 1.74 million years. However, according to models of the post-MS evolution of pop\,III stars, stars of masses above 260\,\msun\ will directly collapse into black holes\footnote{Heger \& Woosley, 2002, ApJ 567, 532.}. This might have generated a pool of numerous massive stellar black holes, yet to be discovered. 

For MS masses in the interval 70 - 260\,\msun, pair-instabilities\footnote{see the review by Woosley, Heger \& Weaver 2002, RevModPhys 74, 1015} lead to violent supernova eruptions (hypernovae) with energies in excess of $10^{51}$\,erg. Only recently has such an energetic {\it hypernova} been reported\footnote {SN2016aps: Nicholl et. al. 2020, arXiv:2004.05840, Nature Astronomy, 13 April 2020; pair-instability supernovae do not produce r-, s- or p-process elements (see Meyer B.S. 1994, ARAA 32, 153), nor species heavier than Zn.} in a metal-poor dwarf galaxy at redshift $z = 0.2657$.

Based on these theoretical results, the absolutely shortest timescale, i.e. for the earliest profound elemental enrichment of the interstellar medium, could conceivably be about only 2 million years after the Big Bang ($z \approx 400$). However, one needs to add the formation time for these heavy pop\,III stars. A rough estimate on that is provided by the free-fall time for gravitational collapse, which would correspond to somewhat less than 5 million years at this time of the evolution of the universe\footnote{Adopting $H_0 = 67.4$\,\kms\,Mpc$^{-1}$ in a standard flat $\Lambda$CDM cosmology, 2 million years after the Big Bang correspond to a redshift $z=420$, temperature $T_{\rm CMB}=1150$\,K and density $n({\rm H})=100$\,\cmthree.}. So, within less than 10 million years ($z \approx 180$), the universe could hypothetically have assembled all the elemental building blocks of the Periodic Table, provided gravitationally unstable density enhancements (``clouds") had formed\footnote{In cosmology, these clouds are called regions of overdensity (see, e.g., Herrera, Waga \& Jor\'as 2017, PhysRevD, 95-6, 4029). In models, an overdensity parameter is introduced, commonly ad hoc with the value $\delta_c = 200$, but other values have also been used. The density of the overdense region is $\rho(t) = \delta_c \times \rho_c$, where $\rho_c = 3/(8 \pi G) \times H^2(t)$ is the critical density. The distribution of the overdense regions is assumed to be Gaussian.}. These clouds would need to have fragmented into a few hundred smaller and denser pieces, corresponding to the seeds of the heavy Pop\,III stars.  

Less than 10 million years would thus constitute the earliest time at which the universe could have been enriched in heavy elements by means of supernova mass ejections, while the less massive SN-progenitors would still be spending time on the main sequence and explode later. Nature's experimenting to generate systems that might eventually become living could have started already then. 
  
 \subsection{The Dark Ages of the universe}
 
 \subsubsection{The standard model}
  
There is, however, a little {\it aber} with the scenario outlined above. Most people are familiar with the fact that gas that is compressed heats up, increasing its pressure build-up and resisting further compression. Something, i.e. some process, would have to get rid of the excess heat, so that the gas could be further compressed\footnote{Isothermal self-gravitating clouds with density contrasts between centre and outer edge $\rho_{\rm 0}/\rho_{\rm out} \ge 32.2$ up to 709 have negative heat capacity (Lynden-Bell D., 1999, Phys A 263, 293, and references therein).}. 
    
 In the standard model, one is following the matter, i.e. the baryonic matter (BM), which, during the period of recombination, only slightly earlier had gone from a completely ionized to an essentially neutral state. At that stage ($z \approx 1100$, age $\approx 470$ thousand years), the temperature of the adiabatically expanding universe is about 3000\,K, with all the hydrogen and helium in their respective electronic ground states. 
 
The CMB consisted of photons with wavelengths around $1\mu$\,m and these hardly interacted with the gas in the clouds. Photons with wavelengths shorter than at least ten times the ones of the CMB would have been required to do anything about this situation. Only these much more energetic photons would have been able to transport the heat away. In summary, it is believed that atomic neutral gas, consisting of hydrogen and helium  (H\,I and He\,I), cannot do the trick. That is, there would not have been luminous things like stars in galaxies illuminating the universe, hence the universe would have appeared dark to the human eye. 
 
 A number of scientists have developed a way out of this dilemma. If one instead of atomic gas uses molecular gas, the heating problem could become largely alleviated. For instance, the hydrogen molecule \molh\ has energy levels that correspond to wavelengths of some micrometers ($\mu$m), i.e. in the energy regime of the CMB at that time. Within an order of magnitude, the formation time\footnote{Gould R.J. \& Salpeter E.E. 1963, ApJ 138, 393; the formation efficiency is believed to be within the range $0.1 - 1$.} of \molh\ is $t({\rm H}_2)$\,\gapprox\,$10^8\,{\rm yr}/(n_{\rm H}$\,\cmthree). This very slow build-up of sufficient amounts of coolants resulted in essentially no radiation of optical photons for extended periods of time, hence the term ``Dark Ages"\footnote{For a review, see Miralda-Escud\'e J. 2003, Sci. 300, 1904.}.

F.\,Palla and collaborators have made detailed models of  \molh\ formation in the early universe, including a reasonably large number of chemical reactions\footnote{Palla F., 1999, in: Proceedings of Star Formation 1999, held in Nagoya, Japan, June 21 - 25, 1999, ed. T. Nakamoto, Nobeyama Radio Observatory, p. 6-11, and references therein. My friend Francesco passed prematurely away in 2016.}. According to these models, an appreciable fraction of \molh\ has formed at $z \approx 300$ [3.3\,Myr, 820\,K, 35\,\cmthree] and reached equlibrium at $z \approx 100$ [17\,Myr, 275\,K, 1\,\cmthree].  These models contained already in their initial set-up molecular ``impurities", as the simplest reaction, i.e. ground state ${\rm H} + {\rm H} \rightarrow {\rm H}_2 + h \nu$, exceeded the age of the universe and would not have been viable\footnote{see, e.g., Latter W.B. \& Black J.H., 1991, ApJ 372, 161. These authors address this reaction by assuming that one of the hydrogen atoms is in an excited state ($\approx 10$\,eV).}.

In the presence of dust, \molh\ formation can be speeded up considerably\footnote{see the review by Wakelam V. et al., 2017, Molecular Astrophysics 9, 1.}.  The afore-mentioned models had not any dust included.   

So, about 20 million years after the Big Bang, the evolution toward unstable self-gravitating entities could have begun. Free-fall times would be some 40 million years, i.e. the first stars could have lightened up within less than 100 million years. However, theoretical models of structure formation need considerably more time to make anything like stars, putting that into the redshift interval $z \approx 20 \,{\rm to}\,10$, which corresponds to 400 million to 1 billion years. In these models, the main actor for the gravitational pulling-together is the dark matter (DM), which does not care much about any cooling radiation, as dark matter does not interact with photons. The ordinary matter (BM) of the becoming stars is thought to have been dragged along, taking a piggy-back ride on the dark matter. 

\subsubsection{How dark were the Dark Ages and for how long?}

Once the first stars had formed, their energetic radiation was re-ionizing the universe. The subsequent recombinations of the protons and electrons gave rise to optical photons, lightening up the universe again. According to the Planck Collaboration\footnote{The Planck collaboration, Planck 2018 results. VI. Cosmological parameters, arXiv:1807.06209v2. For $z=7.7$, I find $T_{\rm CMB}=24$\,K, $n({\rm H})=8.5\,10^{-4}$\,\cmthree\ and $t_{\rm ff}=1.6$\,Gyr.}, the mid-point redshift of this recombination period was $z_{\rm rec} = 7.7 \pm 0.7$, corresponding to the age of $700 \pm 100$ million years\footnote{Harikane Y. et al. (2020, arxivv: 1910.10927v3) discover a large population of galaxies at $z>6$. The redshifts were determined from spectral lines of carbon and oxygen.}.

It has been realized that supernovae can produce copious amounts of dust\footnote{Matsuura et al. 2011, Sci 333, 1258: {\it Herschel Detects a Massive Dust Reservoir in Supernova 1987A}}. If there had been supernova explosions early on in the the universe, the evolution of the matter universe would have been different from what is now the established view. But it is also clear that our understanding of the evolution of the universe, after the decoupling of matter and radiation, depends critically on what the dark matter (DM) was doing, both before and after. Obviously, being ignorant about the true nature of DM, any assumptions concerning its physics must remain rather speculative\footnote{The {\it pro}\hspace{0.025cm}s and {\it con}\hspace{0.025cm}s of current dark matter theories, including the plethora of proposed carriers, some sort of astro-particle, are discussed by Ostriker J.P. \& Steinhardt P. 2003, Sci. 300, 1909, {\it New Light on Dark Matter}. The authors also discuss the concept of Dark Energy, the identity of which is equally unknown, although assumed to be some sort of homogeneous ``fluid". Unlike Dark Matter, which is gravitationally self-attractive, Dark Energy is gravitationally self-repulsive, being the driver of the currently accelerated expansion of the universe.}.

To summarise, we can identify at least two, more or less well-founded, ideas regarding the history of the first stars in the early universe. According to the widely established standard model, the first stars did not form before the universe was almost a billion years old. On the other hand, if the Dark Ages had been very much shorter, or did not really exist at all, the first stars could have formed already 60 million years or so after the Big Bang. These scenarios represent two quite different time scales.

An even shorter time scale has been discussed by A.\,Loeb, according to whom life might have originated already at the age of 10 to 17 million years\footnote{Loeb A. 2014, Int. Journal of Astrobiology 13, 337: {\it The Habitable Epoch of the Early Universe}. At $z + 1 \approx 100-137$, Loeb envisages the existence of rocky planets, whose formation would have required a high degree of non-Gaussianity (\gapprox\,$8\sigma$) in the probability distribution of the initial overdensity regions.}. This interval corresponds to the time during which CMB-temperatures were between $100^{\circ}$ and $0^{\circ}\,{\rm C}$, i.e. the temperatures of water on a rocky planet in the liquid phase. Clearly, the idea of life being of the terrestrial kind has been implicitly assumed. 

The ever improving sensitivity of the observations of remote galaxies, due to larger telescopes and more sophisticated detectors, has widened our horizons since the time of Edwin Hubble to increasingly larger redshifts, which at the moment of writing corresponds to $z>9$. In the context of the present paper the most important point is that the redshift of this galaxy, MACSJ\,1149-JD1, has been determined from the observation of a spectral line of oxygen\footnote{Hashimoto T. et al. (https://almascience.eso.org/alma-science/science-highlights-highz-oxygen) report the discovery of a galaxy at $z=9.1096$. The authors arrive at the conclusion that star formation in that galaxy begun at $z\approx15$, i.e. about 250 million years after the Big Bang.}. This implies that at such high redshifts generations of supernovae had already produced significant amounts of this non-primordial element.

\section{Coming back to the issue of life}
 
One of the prime arguments for the possible ubiquity of life has been that life on Earth uses the universally most abundant elements\footnote{Except helium, which is a noble, i.e. chemically non-reactive, species. Nitrogen is less abundant than neon and iron, being the seventh most abundant element in the Milky Way.} as its basic building blocks, i.e. H, O, C and N: carbon based chemistry in water solution together with nitrogen (for, e.g., the amino acids). It may appear odd therefore that the terrestrial type of life is composed of the least available form of matter, viz. of the ``ordinary" (baryonic) matter contributing merely 4\% to the total energy budget, which is dominated by Dark Energy (73\%) and Dark Matter (23\%).
 
On the other hand, in highly complex structural forms, such as terrestrial DNA, the low-abundance element of phosphorus is incorporated. Like oxygen, that element too required supernova explosions for its production\footnote{Cescutti C., Matteuchi F. Caffau E. \& Franois P. (2012, A\&A 540, A33){\it Chemical evolution of the Milky Way: the origin of phosphorus}}. In the biomolecular literature, there has been a vivid discussion regarding as to what extent important molecules, used by life on Earth, could be replaced by others that are very similar in their functionality. A prime example would be the possible substitution of 15th column phosphorus by dito arsenic\footnote{Knodle R., Agarwal P. \& Brown M. (2012, Biomolecules 2, 282) {\it From Phosphorous to Arsenic: Changing the Classic Paradigm for the Structure of Biomolecules}.}. This has actually been observed in some bacteria on Earth, but, in the end, did not gain the winning hand: phosphates are vastly more exploited by living organisms on Earth than arsenates\footnote{Elias, M., et al., 2012, Nat, 491, 134, {\it The molecular basis of phosphate discrimination in arsenate-rich environments}}. In addition, arsenic is more than a thousand times less abundant than phosphorus; so, in this case, nature actually picked the more plentiful one.

In summary, for arriving at universal beginnings and evolutions of life, large relative quantity seems not to be of primary importance.
 
\section{Possibilities to detect extraterrestrial life} 
 
This chapter will be deliberately kept brief, since there exists a very large body of modern literature to which the reader is referred (see below). In general, that literature covers aspects of astrobiology, biochemistry, geology and the physics of the habitats of exoplanetary and terrestrial life. Here, we will also address some topics that are generally not covered by the conventional literature.
 
 \subsection{Nearby universe}
 
Not having a clear definition of the concept of life makes searching for it difficult, not knowing what to look for. However, one giant thought leap was taken when it was proposed already early on to look for, not simply bacterial but what is called ``intelligent", life with the SETI project\footnote{Visit e.g., https://www.seti.org/seti-institute/Search-Extraterrestrial-Intelligence. See also: Westby \& Conselice (2020, arXiv:2004.03968, {\it The Astrobiological Copernican Weak and Strong}) who attempt to estimate the number of such {\it civilisations} (sic!), where their case is built on a stereotypical parallel to the human history.}. SETI represents the anthropocentric vision that life, given enough time, develops beings similar to ourselves\footnote{But recall the example of the sharks, which for nearly 400 million years haven't changed that much.}. SETI has mostly been listening to, but occasionally also been sending, radio messages at well defined wavelengths. The underlying assumption is that every technically advanced civilisation will know about the 21\,cm (1.4\,GHz) line of hydrogen\footnote{The wavelength of 21\,cm corresponds to the energy difference of the hyperfine states ($F=1$ and $F=0$) of the H\,I 1\,s ground state. This spin flip transition has a lifetime of nearly 11 million years ($A_{10} \approx 3\times 10^{-15}\,{\rm s}^{-1}$), which, because of its implied low absorption probability,  would make it very suitable for interstellar communication. The line was discovered by Ewen H.I. \& Purcell E.M. 1951, Nat. 168, 356: {\it Observation of a Line in the Galactic Radio Spectrum: Radiation from Galactic Hydrogen at 1,420 Mc./sec} }. And use it for interstellar communication. So far, no convincing result has been obtained.

A couple of comments regarding the universality of the 21\,cm line may be in order here. Firstly, our status of technically advanced civilisation is still in its infancy, meaning that these hypothetical others may be more developed and might have been using other channels of communication\footnote{Instead of photons, using e.g. neutrinos, which are travelling close to the speed of light, and having rest masses $\sum m_{0,\,i}\,c^2 < 0.12$\,eV ($m_0 < 2\times 10^{-34}$\,g) for all three flavours $i$ (e-, $\mu$-, $\tau$-neutrino; Planck Consortium 2018). In quantum physics, elementary particles exhibit wavelike behaviour and their wavelength is called the de Broglie wavelength, $\lambda_{\rm dB}=(h/m_0 \upsilon)\,\sqrt{1-(\upsilon/c)^2}$. Here $\lambda$ is the wavelength, $h$ is Planck's constant, $\upsilon$ is the speed of the particle and $c$ that of light, and $m_0$ is the particle's rest mass. The shorter the wavelength (the higher the frequency) the more information can be transmitted. Since neutrinos travel close to the speed of light, the neutrino wavelength would be significantly shorter than 21\,cm. In addition, neutrinos have a very small absorption cross section (Bethe \& Peierls 1934, Nat. 133, 532): it needs a column of water that is 500\,pc long to make sure to catch one.}. This is like comparing smoke signals with a blanket to high speed broad band communication through optical fibres and/or with global satellite systems. Most people are not looking for smoke signals and hence would not be aware that somebody might want to talk to them. Conversely, the smoke signaller would be entirely ignorant of the ever ongoing intercontinental telephone conversations and broadcastings.

In the process of smoke signalling, some puffs might have become distorted, say by the wind, and a message saying ``dinn atf seabnd brin trump" may be difficult to interpret. Which leads to the basic problem of communicating, i.e. to be able to convey an understandable message. Even when it is undistorted, humans have not been able to do so with various species even on their own home planet. In addition and like on Earth, different species may live in separated and completely different environs. Terrestrial examples are humans on land, fish in the sea and worms in the ice. As such, these also have very different behaviours and experiences, making a meaningful ``conversation" less likely. 
 
\subsubsection{Solar system exploration}
 
Contemplated and/or planned observing campaigns include the exploration of the solar system in the hunt for life. In the pipeline are missions to Mars and the Galilean moons of Jupiter. In addition, landings on comets and asteroids are nearly becoming every-day routine. \\
\\
Internet sites to visit include \\
\\
https://solarsystem.nasa.gov/  \\
https://www.esa.int/Enabling\_Support/Operations/Solar\_system \\
https://www.ncbi.nlm.nih.gov/books/NBK540085/ {\it The Search for Life in the Coming Decades, An Astrobiology Strategy for the Search for Life in the Universe}. \\
\\
Top priority is the search for signs of past or present liquid water as a proxy for life (in different physical conditions and for diverse chemistries, methane or ammonia, for instance, could perhaps also do the job). 

\subsubsection{Stars and planets and the ISM of the Milky Way}

Current plans for future observations build on the conditions for terrestrial life and are based on ``what life is doing", i.e. to tracing signs of biotic influence on the planets that life is inhabiting. On Earth, life forms have apparently  transformed the archaic oxygen poor atmosphere to the oxygen rich one of today\footnote{Grenfell J.L. et al. 2010, AsBio, 10, 77: {\it Co-Evolution of Atmospheres, Life, and Climate}. In particular regarding oxygen, see: Meadows V.S. et al. 2018, AsBio, 18, 630, {\it Exoplanet Biosignatures: Understanding Oxygen as a Biosignature in the Context of Its Environment}}.  

In particular, one is aiming at searches for biomarkers toward exoplanetary atmospheres\footnote{Kaltenegger L. 2011, Proc. IAU Symp 280, 302, The molecular universe, eds. J. Cernicharo \& R. Bachiller: {\it Biomarkers of Habitable Worlds: Super-Earths and Earths}.}. Currently, much effort is devoted to the finding of a twin of Earth, i.e. more or less an exact copy of this particular - our - planet. The underlying assumption seems to be that life could develop only in terrestrial conditions throughout the history of the Earth. Specifically taken into consideration are, of course, the abundant existence of liquid water, the obliquity of the planet's rotation axis, the presence of the large moon, the dynamics of surface plate tectonics including volcanism, radioactivity and the interior generation of the global magnetic field and so on.  

In addition, during the history of the Earth, viz. on time scales of billions of years, these conditions had changed profoundly. What regards life, the planet has undergone several documented mass extinctions, when nearly all multicellular organisms had been wiped out\footnote{more than 75\% to 90\%, see, e.g., https://en.wikipedia.org/wiki/Extinction\_event}. In spite of this, new life forms did emerge and continued to strive and developed further. Apparently, in the long run, life is stronger than death.

A considerable number of large molecules have been discovered in the dense interstellar medium\footnote{McGuire, B. A., 2018, ApJS, 239, 17, {\it 2018 Census of Interstellar, Circumstellar, Extragalactic, Protoplanetary Disk, and Exoplanetary Molecules}}, with emphasis on regions that are protected against the harmful ultraviolet radiation from the stars. Of special interest is the detection of prebiotic molecules\footnote{see, e.g, McGuire, B.~A., Carroll, P.~B., \& Garrod, R.~T., 2018, ASPC, 517, 245, {\it Prebiotic Molecules}}$^{ ,}$\footnote{Joergensen, J. K., Favre, C., Bisschop, S. E., Bourke, T. L., van Dishoeck, E. F., \& Schmalzl, M., 2012, ApJ, 757, L 4, Detection of the Simplest Sugar, Glycolaldehyde, in a Solar-type Protostar with ALMA} in the ISM, which eventually may be followed by the finding of macro-biomolecules in the future.

To find life in extra-solar-systems, a number of elaborate technical and inventive solutions have been proposed. These include remote observations of the nearby solar neighbourhood, within a few tens of light-years in particular\footnote{{\it Handbook of Exoplanets},  eds. Deeg H.J. \& Belmonte J.A., Springer, https://doi.org/10.1007/978-3-319-55333-7}. However, without in situ measurements, the risk of false positive results will always be present.

\subsection{The distant universe}

High $z$-experiments regarding the detection of life in the early universe have not been considered in earnest: the technical difficulties aside, we have to remember that any beings of animate material would have lived there billions of years ago. In addition, observing redshift distributions of life would strongly indicate that life originates whenever the conditions are right.  

Indirect, and non-unique, evidence may be obtained from observations of star formation beyond some thresholding redshift, say $z_{\rm ts}$\,\gapprox\,20, when the universe would have been less than 180 million years old\footnote{Bowman et al., 2018, (Nat., 555, 67, {\it An absorption profile centred at 78 megahertz in the sky-averaged spectrum}), attributed an observed broad absorption trough, centred at 78\,MHz and 19\,MHz wide (FWHM), to the 21\,cm line at $20 > z > 15$. The frequency is given by $\nu_{\rm lab} = 1420/(z+1)$\,MHz.}. 

Another, perhaps very remote, possibility could come from a very speculative idea. Assume that an ancient civilisation had reached a high level of understanding of the nature of dark matter. Assume further that DM is, as one generally suspects, made of highly non-interacting heavy particles and that this civilisation had learned how to produce and to control these. Speculating that the DM beams would be kept better collimated over long distances than laser beams, they might have been used for cosmic communication.  We would not be aware of it\footnote{According to Vega, Salucci \& Sanchez (2012, New Astronomy 17-7, 653), cold DM particles have a rest mass in the range 1 to 2\,keV. Being by definition non-relativistic (``cold" means ``slow"), they travel at speeds below 10\% of the speed of light, and their de Broglie wavelength would be $\lambda_{\rm dB}$\,\gapprox\,80\,\AA. The extremely different mass scale of $m_a \approx 10^{-21}$\,eV was considered by Marsh \& Silk (2014, MNRAS, 437, 2652, {\it A model for halo formation with axion mixed dark matter}), which illustrates the difficulty to identify the nature of dark matter.}. 

Because of their exotic status and our ignorance regarding dark matter and dark energy\footnote{Dark energy (DE) comprises 68\% of all the energy density in the universe, with normal matter (BM) contributing just 5\%, with the rest being DM. Dark energy is introduced by $\Lambda$ in Einstein's field equations that counteracts gravity. For some inexplicable reason, $\Lambda$ began to dominate at $z \approx 0.7$, about 6.5 billion years ago. Why?} these speculations may provide yet another set of solutions to the Fermi paradox\footnote{See, e.g., https://en.wikipedia.org/wiki/Fermi\_paradox    ``where is everybody?"}.

\section{Conclusions}
 
 In an attempt to understand the origin(s) of life in the history of the universe, I have been quite naively concentrating on what might have been possible and what not. I tried to avoid as much as possible to lean too much on our experience of life on Earth. Similarly, I tried not to favour (or to disfavour) any particular cosmological model. The one I have used throughout, I picked essentially for calibration purposes: anyone can recalculate the given results according to the model of his or her liking\footnote{On a cosmological clock, Big Bang occurs at $t=0$ and times refer to ``standard" $\Lambda$CDM cosmology with $H_0=67.4$\,\kms\,Mpc$^{-1}$ ($\rho_{\rm c,\,0}=8.53 \times  10^{-30}$\,g\,\cmthree), $\Omega_0 = \Omega_{\Lambda,\,0} + \Omega_{\rm m,\,0} = 1$, $ \Omega_{\rm m,\,0}  = 0.31$, $\Omega_{\kappa} = 0$.}. 
  
Broadly speaking, one could imagine three scenarios for the first emergence of life in the universe: (A) a very early epoch, (B) an early epoch, (C) an archaic terrestrial epoch. Requiring baryonic building blocks, these can briefly be summarised as \\ 

 \noindent
{\bf (A)} a very early epoch, age of the universe less than 0.1 billion years. Life originated at the very first instance it had a chance to do so. Structure formation occurred in conditions of high non-Gaussianity and the formation of life progressed after recombination. The duration of ``darkness" was very brief. Tests of this hypothesis would include evidence for the very early existence of massive Pop\,III stars, at redshifts $z \gg 10$ (e.g. detection in distant galaxies of ${\rm Ly}\,{\alpha}$ at wavelengths considerably longer than 1.3\,\um\ with a Giant Infrared Space Telescope Array); \\

 \noindent
{\bf (B)} an early epoch, age of the universe about 1 billion years. The basis for this case is the widely adopted standard cosmology. In that model, life might have originated after re-ionisation, which is defining the end of the Dark Ages. Again, with regard to life, tests would be indirect and should be able to provide evidence that there are no stars in galaxies at redshifts much larger than six ($z \ll 10$);\\

 \noindent
{\bf (C)} an archaic terrestrial epoch, age of the universe approaching 10 billion years. Life originated first on Earth, once the planet had cooled to sustainable temperatures (about one billion years after formation). This is the opinion shared by many, but could be disproved by finding evidence for life elsewhere, either as relics or as signs of present activity.  The possibility of panspermia exchange within the solar system could prohibit a conclusive answer, unless molecular chiralities are different in different locations. Another, more stringent, hypothesis disproval would therefore have to come from extrasolar environments. An idea is to use biomarkers and to look for chemistry out of equilibrium\footnote{D. Defr\`ere, D., \& L\'eger, A., et al., 2018, Experimental Astronomy 46, 543, {\it Space-based infrared interferometry to study exoplanetary atmospheres}}. Maybe, it could be shown that life is the most probable driver and maintainer of the non-equilibrium. This constitutes, in fact, the basis of recent proposals for large telescopes/arrays in space of the relatively near future.\\

\section{Resum\'e}

In the end, I was not able to provide a unique and definite answer to the question ``when did living organisms first emerge?" A couple of suggestions have been offered, none of which might be true. These differ by a factor of a hundred in time and are based on the conventional assumption that life needs energy sources that generally take the form of stars. And that life if thriving in the surrounding of these stars. The stars are made of baryonic matter. 

Only recently were we caught by complete surprise, when we had to realise that there is something else than ``the matter as we know it". And this something, dark matter and dark energy, is ruling the universe. What will be the next big surprise?

\newpage
\vspace*{1.7cm}
\noindent
{\bf Bibliography} \\
\\
\footnotesize
Alpher, R. A., Bethe, H., \& Gamow, G., 1948, Phys. Rev., 73, 803, {\it The Origin of Chemical\\
\hspace*{1.5cm}Elements} \\
Baraffe, I., Heger, A., \& Woosley, S.~E., 2001, ApJ, 550, 890, {\it On the Stability of Very Massive Primordial\\
\hspace*{1.5cm}Stars}\\
Bernal, J. L., Verde, L. \& Riess, A. G., 2016, Journal of Cosmology and Astroparticle  Physics, 10, {\it The\\
\hspace*{1.5cm}trouble with H$_{0}$} \\
Bethe, H., \& Peierls, R., 1934, Nat., 133, 532, {\it The ``Neutrino"} \\
Bich, L. \& Green, S., 2018, Synthese, 195, 3919, {\it Is defining life pointless? Operational definitions at the\\
\hspace*{1.5cm}frontiers of biology} \\
Bowman, J. D.,  Rogers, A. E.~E., Monsalve, R. A., Mozdzen, T. J. \& Mahesh, N., 2018, Nat., 555, 67,\\
\hspace*{1.5cm} {\it  An absorption profile centred at 78 megahertz in the sky-averaged spectrum} \\
Cescutti, G., Matteucci, F., Caffau, E. \& Fran{\c{c}}ois, P., 2012, A\&, 540, A33, {\it Chemical evolution of the\\
\hspace*{1.5cm}Milky Way: the origin of phosphorus} \\
Cooke, R. J. \& Fumagalli, M., 2018, Nat. Astron., 2, {\it Measurement of the primordial helium abundance\\
\hspace*{1.5cm}from the intergalactic medium} \\
Deeg, H. J. \& Belmonte, J. A., 2018, {\it Handbook of Exoplanets} \\
{Defr{\`e}re}, D., {L{\'e}ger}, A.,  {Absil}, O.,  {Beichman}, C., {Biller}, B., {Danchi}, W.~C., {Ergenzinger}, K., {Eiroa}, C., \\
\hspace*{1.5cm}  {Mu{\~n}oz}, A., Garc{\'\i}a, {Gillon}, M.,  {Glasse}, A., {Godolt}, M. , {Ertel}, S.,  {Fridlund}, M., {Grenfell}, J.~L.,\\
\hspace*{1.5cm}{Kraus}, S., {Labadie}, L., {Lacour}, S., {Liseau}, R., {Martin}, G., {Mennesson}, B., {Micela}, G., {Minardi}, \\
\hspace*{1.5cm}S., {Quanz}, S.~P.,  {Rauer}, H., {Rinehart}, S., {Santos}, N.~C., {Selsis}, F., {Surdej}, J., {Tian}, F., {Villaver},\\ 
\hspace*{1.5cm}E., {Wheatley}, P.~J.,  {Wyatt}, M., 2018, Experimental Astronomy, 45, 543, {\it Space-based infrared\\
\hspace*{1.5cm}interferometry to study exoplanetary atmospheres} \\
Dicke, R. H., Peebles, P. J. E., Roll, P. G. \& Wilkinson, D. T., 1965, ApJ, 142, 414, {\it Cosmic Black-Body\\
\hspace*{1.5cm}Radiation.} \\
Di Valentino, E., Melchiorri, A., \& Silk, J., 2020, Nat. Astronomy, 4, 196, {\it Planck evidence for a closed\\
\hspace*{1.5cm}Universe and a possible crisis for cosmology}\\
Elias, M., Wellner, Al., Goldin-Azulay, K., Chabriere, E., Vorholt, J. A., Erb, T. J., Tawfik, D. S., 2012,  \\
\hspace*{1.5cm}Nat., 491, 134, {\it The molecular basis of phosphate discrimination in arsenate-rich environments}\\
Ewen, H. I. \& Purcell, E. M., 1951, Nat., 168, 356, {\it Observation of a Line in the Galactic Radio Spectrum:\\
\hspace*{1.5cm}Radiation from Galactic Hydrogen at 1,420 Mc./sec.} \\
Gould, R. J. \& Salpeter, E. E., 1963, ApJ, 138, 393, {\it The Interstellar Abundance of the Hydrogen Molecule.\\
\hspace*{1.5cm}I. Basic Processes.} \\
Grenfell, J. L., Rauer, H., Selsis, F., Kaltenegger, L., Beichman, C., Danchi, W., Eiroa, C., Fridlund, M.,\\
\hspace*{1.5cm}Henning, T., Herbst, T., Lammer, H., L{\'e}ger, A., Liseau, R., Lunine, J., Paresce, F., Penny, A.,\\
\hspace*{1.5cm}Quirrenbach, A., R{\"o}ttgering, H., Schneider, J., Stam, D., Tinetti, G. \& White, G. J., 2010,\\
\hspace*{1.5cm}Asbio, 1, 77, {\it Co-Evolution of Atmospheres, Life, and Climate} \\
Harikane, Y., Laporte, Ni., Ellis, R. S. \& Matsuoka, Y., 2020, arXiv:2005.11078, {\it The Mean Absorption\\
\hspace*{1.5cm}Line Spectra of a Selection of Luminous z\raisebox{-0.5ex}\textasciitilde6 Lyman Break Galaxies} \\
Hashimoto, T. 2020, IAU Symp., 352, 13, {\it Properties of galaxies at z $\approx 7 - 9$ revealed by ALMA} \\
Heger, A., Woosley, S. E., 2002, ApJ, 567, 532, {\it The Nucleosynthetic Signature of Population III}\\
Herrera, D., Waga, I., \& Jor\'as, S. E., 2017, Phys. Rev. D 95-6, 4028 {\it Calculation of the critical over-\\ 
\hspace*{1.5cm}density in spherical-collapse approximation} \\
Howes, L. M., Casey, A. R., Asplund, M.,Keller, S. C., Yong, D., Nataf, D. M., Poleski, R., Lind, K.,\\
\hspace*{1.5cm}Kobayashi, C., Owen, C. I., Ness, M., Bessell, M. S., da Costa, G. S., Schmidt, B. P., Tisserand,\\
\hspace*{1.5cm}P.,Udalski, A., Szyma{\'n}ski, M. K., Soszy{\'n}ski, I., Pietrzy{\'n}ski, G., Ulaczyk, K., Wyrzykowski, {\L}.,\\
\hspace*{1.5cm}Pietrukowicz, P., Skowron, J., Koz{\l}owski, S. \& Mr{\'o}z, P., 2015, Nat., 527, 484, {\it Extremely metal-\\
\hspace*{1.5cm}poor stars from the cosmic dawn in the bulge of the Milky Way} \\
Joergensen, J. K., {Favre}, C., {Bisschop}, S. E., {Bourke}, T. L.,  {van Dishoeck}, E. F., \& {Schmalzl}, M., 2012,\\
\hspace*{1.5cm} ApJ, 757, L\,4,  {\it Detection of the Simplest Sugar, Glycolaldehyde, in a Solar-type Protostar with\\
\hspace*{1.5cm}ALMA} \\
Kaltenegger, L., 2011, IAU Symp., 280, 302, {\it Biomarkers of Habitable Worlds - Super-Earths and Earths} \\
Knodle, R., Agarwal, P., Brown, M., 2012, Biomolecules, 2, 282, {\it From Phosphorous to Arsenic: Changing\\
\hspace*{1.5cm}the Classic Paradigm for the Structure of Biomolecules} \\
Latter, W. B. \& Black, J. H., 1991, ApJ, 372, 161, {\it Molecular Hydrogen Formation by Excited Atom\\
\hspace*{1.5cm}Radiative Association} \\
Loeb, A., 2014, International Journal of Astrobiology, 13, 337, {\it The habitable epoch of the early Universe} \\
Lough, W. J., \& Wainer, I. W., eds., 2002, {\it Chirality in natural and applied science}, Oxford: Blackwell \\
\hspace*{1.5cm}Science \\
Lynden-Bell, D., 1999, Physica A Statistical Mechanics and its Applications, 263, 293, {\it Negative Specific\\
\hspace*{1.5cm}Heat in Astronomy, Physics and Chemistry} \\
Marsh, D. J. E., \& Silk, J., 2014, MNRAS, 437, 2652, {\it A model for halo formation with axion mixed dark \\
\hspace*{1.5cm} matter} \\
Matsuura, M., Dwek, E., Meixner, M., Otsuka, M., Babler, B., Barlow, M. J., Roman-Duval, J.,\\
\hspace*{1.5cm}Engelbracht, C., Sandstrom, K., Laki{\'c}evi{\'c}, M., van Loon, J. Th., Sonneborn, G., Clayton,\\
\hspace*{1.5cm}G.C., Long, K. S., Lundqvist, P., Nozawa, T., Gordon, K. D., Hony, S., Panuzzo, P., Okumura\\ 
\hspace*{1.5cm}K., Misselt, K. A., Montiel, E., \& Sauvage, M., 2011, Sci., 333, 1258, {\it Herschel Detects a\\
\hspace*{1.5cm}Massive Dust Reservoir in Supernova 1987A} \\
McGuire, B. A., 2018, ApJS, 239, 17, {\it 2018 Census of Interstellar, Circumstellar, Extragalactic, Proto-\\
\hspace*{1.5cm}planetary  Disk, and Exoplanetary Molecules} \\
McGuire, B.~A., Carroll, P.~B., \& Garrod, R.~T., 2018, ASPC, 517, 245, {\it Prebiotic Molecules} \\
Meadows, V. S., Reinhard, C. T., Arney, G. N., Parenteau, M. N., Schwieterman, E. W., Domagal-\\
\hspace*{1.5cm}Goldman, S. D., Lincowski, A. P., Stapelfeldt, K. R., Rauer, H., DasSarma, S., Hegde, S.,\\ 
\hspace*{1.5cm}Narita, N., Deitrick, R., Lustig-Yaeger, J., Lyons, T. W., Siegler, N., Grenfell, J. L., 2018, \\
\hspace*{1.5cm}Asbio, 18, 630, {\it Exoplanet Biosignatures: Understanding Oxygen as a Biosignature in  \\
\hspace*{1.5cm}the Context of Its Environment} \\
Meyer, Bradley S., 1994, ARAA, 32, 153, {\it The r-, s-, and p-Processes in Nucleosynthesis} \\
Miralda-Escud{\'e}, J., 2003, Sci., 300, 1904, {\it The Dark Age of the Universe} \\
Nicholl, M., Blanchard, P. K., Berger, E., Chornock, R., Margutti, R., Gomez, S., Lunnan, R., Miller,\\
\hspace*{1.5cm}A. A., Fong, W., Terreran, G., Vigna-G{\'o}mez, A., Bhirombhakdi, K, Bieryla, Al., Challis, P.,\\
\hspace*{1.5cm}Laher, R.R., Masci, Frank, J. \& Paterson, K., 2020, Nat. Astronomy, arXiv2004.05840,\\
\hspace*{1.5cm}{\it An extremely energetic supernova from a very massive star in a dense medium} \\
Nordlander, T., Bessell, M.~S., Da Costa, G.~S., Mackey, A.~D., Asplund, M., Casey, A.~R., Chiti, A.,\\ 
\hspace*{1.5cm}Ezzeddine, R., Frebel, A., Lind, K.,Marino, A.~F., Murphy, S.~J., Norris, J.~E., Schmidt, B.~P.\\
\hspace*{1.5cm}\& Yong, D., 2019, MNRAS, 488, L109, {\it The lowest detected stellar Fe abundance: the halo star\\
\hspace*{1.5cm}SMSS J160540.18-144323.1}\\
Oelert, W., for the PS210-Collaboration, 1997, Nuc.Phys. B, Proceedings Suppl., 56, 319, {\it Production of\\
\hspace*{1.5cm}anti-hydrogen at CERN}\\
Ostriker, J. P. \& Steinhardt, P., 2003, Sci., 300, 1909, {\it New Light on Dark Matter}\\
Palla, F., 1999, Proc. Star Formation 1999, 6, ed. Nakamoto, T., {\it The Formation of Primordial Stars}\\
Penzias, A.~A. \& Wilson, R.~W., 1965, ApJ, 142, 419, {\it A Measurement of Excess Antenna Temperature\\ 
\hspace*{1.5cm}at 4080 Mc/s.}\\
Planck Collaboration, 2018, arXiv:1807.06209, {\it Planck 2018 results. VI. Cosmological parameters} \\
Riess, A. G., Casertano, S. Yuan, W., Macri, L. M. \& Scolnic, D., 2018, ApJ, 876, 85, {\it Large Magellanic\\ 
\hspace*{1.5cm}Cloud Cepheid Standard Candles Provide a 1\% Foundation for the Determination of the Hubble\\
\hspace*{1.5cm}Constant and Stronger Evidence for Physica beyond $\Lambda$CDM}\\
Riess, A. G., 2019, Nat. Rev. Phys., 2, 10, {\it The expansion of the Universe is faster than expected}\\
Smith, M., Sullivan, M., Nichol, R.~C., Galbany, L., D'Andrea, C.~B., Inserra, C., Lidman, C., Rest, A.,\\
\hspace*{1.5cm}Schirmer, M.,Filippenko, A.~V., Zheng, W., Cenko, S. Bradley,Angus, C.~R., Brown, P.~J., \\
\hspace*{1.5cm}Davis, T.~M., Finley, D.~A., Foley, R.~J., Gonz{\'a}lez-Gait{\'a}n, S., Guti{\'e}rrez, C.~P., Kessler, R.,\\
\hspace*{1.5cm}Kuhlmann, S., Marriner, J., M{\"o}ller, A., Nugent, P.~E., Prajs, S., Thomas, R., Wolf, R.,\\
\hspace*{1.5cm}Zenteno, A., Abbott, T.~M.~C., Abdalla, F.~B., Allam, S., Annis, J., Bechtol, K., Benoit-L{\'e}vy, \\
\hspace*{1.5cm}A., Bertin, E., Brooks, D., Burke, D.~L., Carnero, Rosell., A., Carrasco Kind, M., Carretero, \\
\hspace*{1.5cm}J., Castander, F.~J., Crocce, M., Cunha, C.~E., da Costa, L.~N., Davis, C., Desai, S., Diehl, \\
\hspace*{1.5cm}H.~T., Doel, P., Eifler, T.~F., Flaugher, B., Fosalba, P., Frieman, J., Garc{\'\i}a-Bellido, J. \\
\hspace*{1.5cm}Gaztanaga, E., Gerdes, D.~W., Goldstein, D.~A., Gruen, D., Gruendl, R.~A., Gschwend, J., \\
\hspace*{1.5cm}Gutierrez, G., Honscheid, K., James, D.~J. , Johnson, M.~W.~G., Kuehn, K., Kuropatkin, N., \\
\hspace*{1.5cm}Li, T.~S., Lima, M., Maia, M.~A.~G., Marshall, J.~L., Martini, P., Menanteau, F., Miller, C.~J.,\\
\hspace*{1.5cm}Miquel, R., Ogando, R.~L.~C., Petravick, D., Plazas, A.~A., Romer, A.~K., Rykoff, E.~S., Sako, \\
\hspace*{1.5cm}M., Sanchez, E., Scarpine, V., Schindler, R., Schubnell, M., Sevilla-Noarbe, I., Smith, R.~C., \\
\hspace*{1.5cm}Soares-Santos, M., Sobreira, F., Suchyta, E., Swanson, M.~E.~C., Tarle, G., Walker, A.~R. and \\
\hspace*{1.5cm}DES Collaboration, 2018, ApJ, 854, 37, {\it Studying the Ultraviolet Spectrum of the First Spectro-\\
\hspace*{1.5cm}scopically Confirmed Supernova at Redshift Two} \\
Thomas, R.~C. \& Kantowski, R., 2000, Phys.Rev. D 62, 10, {\it Age-redshift relation for standard cosmology} \\
Vega, H. J., Salucci P., \& Sanchez, N. G., 2012, New Astronomy, 17-7, 653, {\it The mass of the dark matter \\
\hspace*{1.5cm}particle: Theory and galaxy observations} \\
Wakelam, V., Bron, E., Cazaux, S., Dulieu, F., Gry, C., Guillard, P., Habart, E., Hornek{\ae}r, L.,Morisset, \\
\hspace*{1.5cm}S., Nyman, G., Pirronello, V., Price, S. D., Valdivia, V., Vidali, G. \& Watanabe, N., 2017, \\
\hspace*{1.5cm}Molecular Astrophysics, 9, 1, {\it H$_{2}$ formation on interstellar dust grains: The viewpoints of the- \\
\hspace*{1.5cm}ory, experiments, models and observations} \\
Westby, T., \& Conselice, C. J., 2020, arXiv:2004.03968, {\it The Astrobiological Copernican Weak and Strong\\
\hspace*{1.5cm} Limits for Extraterrestrial Intelligent Life} \\
Wetherill, G.~W., 1975, Annual Review of Nuclear and Particle Science, 25, 283, {\it Radiometric Chronology\\
\hspace*{1.5cm}of the Early Solar System} \\
Wong, K. C., Suyu, S. H., Chen, G. C. -F., Rusu, C. E., Millon, M., Sluse, D., Bonvin, V., Fassnacht, C.  \\
\hspace*{1.5cm}D., Taubenberger, S., Auger, M. W., Birrer, S. Chan, J. H.~H., Courbin, F., Hilbert, S., Tih-\\
\hspace*{1.5cm}honova, O., Treu, T., Agnello, A., Ding, X., Jee, I., Komatsu, E., Shajib, A. J., Sonnenfeld, A., \\
\hspace*{1.5cm}Blandford, R. D., Koopmans, L. V.~E., Marshall, P. J. \& Meylan, G., 2020, MNRAS, arXiv1907.04869, \\
\hspace*{1.5cm}{\it H0LiCOW XIII. A 2.4\% measurement of H$_{0}$ from lensed quasars: 5.3$\sigma$ tension between early \\
\hspace*{1.5cm}and late-Universe probes} \\
Woosley, S.~E., Heger, A. \& Weaver, T.~A., 2002, Rev.Mod.Phys., 74, 1015, {\it The evolution and\\
\hspace*{1.5cm}explosion of massive stars}\\          

\newpage

\section*{Appendix}

\noindent
A rough time line for events that might have had relevance for the development of life in the universe. The numbers refer to the age of the universe according to the cosmology referred to in the text. \\

\noindent
{\bf 370 thousand years ($z=1300$)}: the ionized plasma recombines into neutral matter. Beginning of the Dark Ages (standard theory).

\noindent
{\bf 470 thousand years ($z=1100$)}: is as far back as we are able to observe the universe by means of the cosmic microwave radiation (CMB). Ordinary (baryonic) matter (BM) has already been more important than radiation since the age of 40 thousand years. 

\noindent
{\bf 7 million years ($z=180$)}: is the age at which the very first stars might have formed (H\,I + DM). These would have been massive (several hundred times the mass of the sun) and exploded within less than 2 million years after their birth as hypernovae, sheding highly enriched material into the surrounding medium.

\noindent
{\bf 10 million years ($z=136$)}: dark matter (DM) begins to become more important than BM. The CMB temperature is 100\adeg\,C.

\noindent
{\bf  17 million years ($z=99$)}: molecular hydrogen has formed, allowing the cooling of the gas through vibrationally excited \molh. The CMB temperature is 0\adeg\,C.

\noindent
{\bf 60 million years ($z=45$)}: possibly, formation of the first stars (with \molh). Their UV radiation would have re-ionised the universe in an early, non-standard, ending of the Dark Ages.

\noindent
{\bf 700 million years ($z=7.7$)}: midpoint, within 100 million years, of the Dark Ages, according to the standard model ($\Lambda$CDM cosmology).

\noindent
{\bf 900 million years ($z=6$)}: the end of the Dark Ages, with the first stars re-ionising the universe (standard theory).

\noindent
{\bf 7.3 billion years ($z=0.7$)}: dark matter (DM) begins to become more important than ``ordinary" matter (BM).

\noindent
{\bf 9.3 billion years ($z=0.415$)}: formation of the solar system and the birth of Earth.

\noindent
{\bf 13.8 billion years ($z=0.0$)}: humans write the history of the universe, as they understand it (or not).

\end{document}